# A HYBRID APPROACH TOWARDS INTRUSION DETECTION BASED ON ARTIFICIAL IMMUNE SYSTEM AND SOFT COMPUTING

**Sugata Sanyal**
School of Technology and Computer Science,
Tata Institute of Fundamental Research, Mumbai, India
sanyals@gmail.com

**Manoj Rameshchandra Thakur\***
Computer Science Department, VJTI, Mumbai, India
manoj.thakur66@gmail.com

\*Corresponding Author

### Abstract

A number of works in the field of intrusion detection have been based on Artificial Immune System and Soft Computing. Artificial Immune System based approaches attempt to leverage the adaptability, error tolerance, self-monitoring and distributed nature of Human Immune Systems. Whereas Soft Computing based approaches are instrumental in developing fuzzy rule based systems for detecting intrusions. They are computationally intensive and apply machine learning (both supervised and un-supervised) techniques to detect intrusions in a given system. A combination of these two approaches could provide significant advantages for intrusion detection. In this paper we attempt to leverage the adaptability of Artificial Immune System and the computation intensive nature of Soft Computing to develop a system that can effectively detect intrusions in a given network.

## 1. INTRODUCTION

### 1.1 Background

The average volume of internet traffic has increased many-fold in recent times. Some primary reasons for the increase in the internet traffic are: advancement of the cloud infrastructure for hosting service, emergence of a number of social networking websites and tremendous increase in the use of mobile platform for accessing services hosted on the web [8][9]. With such increase in the volumes of internet traffic, security has become a major concern. Data loss, malicious usage of computing resources and illegal data access are some of the problems that enterprise systems suffer from [2]. There exists a need for intelligent and adaptive systems that can not only detect malicious activities in the network but also evolve over time to mitigate the effects of ever increasing security threats and attacks.

The similarity between an intrusion detection system and the human immune system is quite remarkable. The primary function of the human immune system is to identify malicious external agents like viruses and pathogens, from the native healthy tissue and cells. The sample space with which the Human Immune System works is huge since the human body interacts with innumerable external agents. The Human Immune system also evolves over time to the tackle the ever evolving pathogens and other malicious external agents. The Human Immune

System employs a layered defense approach with the specificity of the defenses increasing with each level. Like the Human Immune System an Intrusion Detection System is also expected to defend the network or the distributed computing infrastructure from malicious attacks like DDoS attack, key-logging attack and Ping of Death attack to name a few. An Intrusion Detection System is also expected to evolve over time to combat new attacks that emerge over time. Thus an Intrusion Detection System design, based on a Human Immune System can provide significant advantages. Like the Human Immune System, the Intrusion Detection System has to work with a large amount of data (network traffic traces). With such high volume of data, it is extremely difficult to predict exactly the occurrence of an intrusion. This is primarily because the malicious activities (traffic data) form a very small portion of the sample space (internet traffic) and the pattern it follows varies based on the type of intrusion or attack that the system is suffering from. Moreover there exists a high level of uncertainty in the sample data; as a result it is impossible to predict the occurrence of an intrusion with a probability of 1. Due to such high volumes and the existence of uncertainty in the data, a technique that is designed along the lines on the Human Immune System but employs Soft Computing techniques at each defense layer can be effective in detecting an intrusion in a given system. In this paper we propose a system based on the above mentioned technique.

### 1.2 Related Work

*Artificial Immune System*

A number of works related to intrusion detection have been inspired by the human immune system. [20] presents one of the first lightweight intrusion detection systems based on AIS (Artificial Immune System) [19]. [21] presents an intrusion detection system based on the emerging 'Danger Theory' [30]. Previously suggested Intrusion detection Systems based on AIS models have used one of the following algorithms; negative selection algorithm, clonal selection algorithm, artificial immune network, danger theory inspired algorithms and dendritic cell algorithms [32]. [22] attempts to solve the problems with contemporary intrusion detection systems using autonomous agents. [13] is based on AIS and attempts to detect abnormality in electromagnetic signals in a complex electromagnetic environment.[1] presents an analogy between the human immune system and the intrusion detection system. It is important to note that this work attempts to evolve the Primary Immune Responses to a Secondary Immune Response using genetic operators like selection, cloning, crossover and mutation. [23] focuses on specifically *static clonal selection* with a negative selection operator. [38] describes a genetic classifier-based intrusion detection system.

*Soft Computing and Machine Learning:*

The lack of exactness and inconsistency in the network traffic patterns has encouraged a number of attempts towards intrusion detection based on 'Soft Computing' [11] [12]. 'Soft Computing' techniques attempt to devise inexact and approximate solutions to the computationally-hard task of detecting abnormal patterns corresponding to an intrusion. [4] proposes a Soft Computing based approach towards intrusion detection using a fuzzy rule based system. [15] suggests an approach based on machine learning techniques for intrusion detection. [37] applies a combination of protocol analysis and pattern matching approach for intrusion detection.[18] proposes an approach towards intrusion detection by analyzing the system activity for similarity with the normal flow of system activities using classification trees. [10] presents a proactive detection and prevention technique for intrusions in a Mobile Ad hoc Networks (MANET).

The rest of the paper is structured as follows: Section 2 introduces the design of the proposed system, followed by detailed explanation of each of the components of the system in sections 3, 4 & 5. Section 6 explains the techniques used for secure propagation of information to other instances of the system in a typical network deployment followed by the conclusion in Section 7.

## 2. THE DESIGN

The design of the proposed system is inspired by the Human Immune System; as a result the components of the system have one-one correspondence with the components of the Human Immune System. The system is divided into three primary components:

1) Surface Barrier
2) Innate Immune system
3) Adaptive system

The network traffic entering the system has to pass through each of the three components. Figure.1 shows orientation of the three components and the flow of network traffic through the three components.

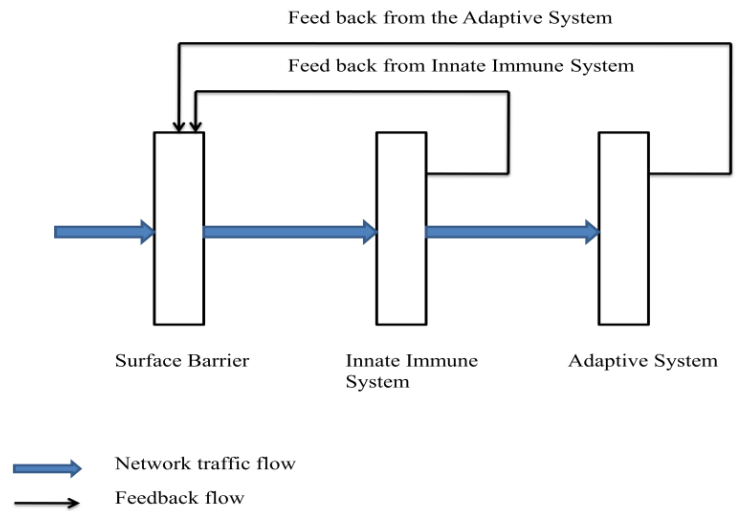

**Figure1 Three primary components of the system**

For each component we define a probability value *P* which is the probability with which the component detects an intrusion in the system. The value for *P* is determined and improved based on the training of the system as well as the feedback from the previous executions of the system. The system thus exhibits a self-improving adaptive nature. The proposed system goes through the following three phases:

1) Training Phase: during this phase each of the three components of the system are trained using training data set. Results of the training phase enable each of the three components to detect abnormalities and hence intrusion in live production network traffic data.
2) Detection Phase: this phase involves the actual detection of intrusion in the system. The system encounters live network traffic in this phase.
3) Feedback Phase: data encountered during the detection phase is fed back to the system to improve the performance and efficiency of the components of the system. Even though the generation of the feedback to improve the efficiency of the system is mentioned as a separate phase, the generation and assimilation of the feedback data into the system is a continuous process.

There exists an instance of the system running on every host of a given network. Information deduced during the training and the detection phase by these instances is broadcasted to other instances of the system to ensure that all the instances are in sync with each other. Figure 2 shows a diagrammatic representation of the instances of the proposed system in a typical network deployment. A detailed explanation of the techniques used for secure transmission of information across the network to other instances of the system is presented in Section 6.

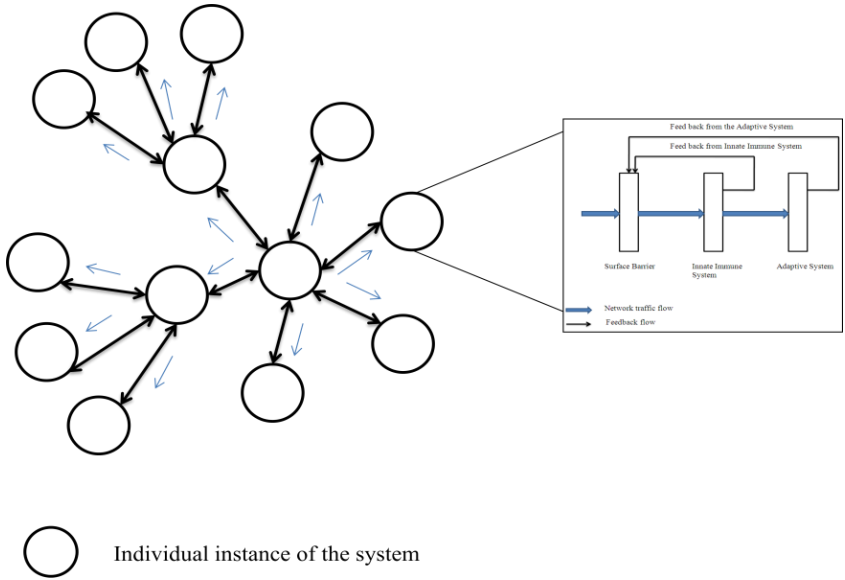

Figure2 Network deployment of the instances of the system

Succeeding sections explain the functioning of the three components in each of the three phases mentioned above.

## 3. SURFACE BARRIER

Surface barriers constitute the first line of defense of the system. This component is analogous to the human skin which acts as the first line of defense against infection. The detectors in this component perform on absolute detection. The term 'absolute detection' indicates that the component detects intrusions based on precise and crisp values, the concept is similar to hard computing wherein the rules defined can detect occurrences with absolute certainty. Thus the value for $P$ in case of surface barrier is 1. Surface barrier uses the following set of information for detecting intrusions at a primary level:

1) IP addresses of malicious hosts in the network
2) Malicious TCP ports used for performing standard attacks like DDoS, Ping of Death, etc. For example IRC bots use port 6667 to perform attacks based on the commands issued by the bot master.

It must be noted that surface barrier doesn't possess any intelligence or computational capability; it attempts to detect intrusions at a primary level merely based on value matching operation.

Surface barrier starts with an empty set of data that it uses to match the parameter values for network traffic data. Over a period of multiple executions of the system in the training phases, surface barrier builds an extensive data set consisting of values for parameters like IP addresses and TCP ports from the feedback from the other two components in the training phase itself. The standard set of data against which surface barrier operates is thus derived initially from the training data and externally fed information and then grows over a period of time as the system encounters more network traffic.

## 4. INNATE IMMUNE SYSTEM

Network traffic that successfully passes through the surface barriers encounter the second line of defense called the Innate Immune System. The Innate Immune System attempts to detect intrusions in the system based on pattern recognition. This component is superior to the surface barriers because it possesses the capability and computational power for performing pattern recognition. The responses issued by Innate Immune System based on the results of the pattern recognition task and the responses initiated by the Surface Barrier form the Primary Immune Response. It is important to note that the each Innate Immune System detector operate on the stream of network traffic data based on a specific characteristics of the network traffic. Thus each detector is characterized by a specific network parameter *C*, where *C* could be incoming data rate in bytes/sec, outgoing data rate in bytes/sec, time required to handle incoming request etc. The patterns inferred during the training phase and the previous executions of the system are added to the pattern set of the individual detectors appropriately based on the network traffic parameter that characterizes the detector. For example, a pattern inferred, of the incoming network traffic in bytes/sec is added to the pattern set of the receptor that characterizes the incoming network data traffic.

Innate Immune System starts with some standard self and non-self patterns fed to the system externally. Over multiple executions in the training phase it develops new self and non-self patterns that are used during the detection phase. Data encountered during the detection phase is fed back to the system to improve the already inferred self and non-self patterns. A detailed explanation of the pattern recognition process is as follows:

**4.1 Pattern analysis based on Dynamic Time Warping**

The detectors in Innate Immune System use Dynamic Time Warping (DTW) algorithm; often used in speech recognition techniques, for pattern matching. [45] presents a distributed approach for detecting botnets in an enterprise network. It uses DTW algorithm to detect abnormal flow patterns. Use of DTW algorithm by the Innate Immune System for pattern recognition is on similar lines as suggested in [45].Each detector tries to match the network traffic data stream with each of the patterns in its pattern set using DTW. DTW is an algorithm that measures the similarity between two sequences which may vary in time or speed. The detectors use DTW algorithm primarily because:

• The sequences that can be compared using this algorithm may vary in time and speed.
• The algorithm provides a non-linear (elastic) alignment, which produces a more intuitive similarity measure, allowing similar network traffic patterns to match even if they are out of phase across the time axis [24].
• The kind of sequences that DTW algorithm can analyze is similar to network traffic sequences that the Innate Immune System detectors analyze.

To calculate the similarity index for two sequences A, B having *m*, *n* data points, respectively, the following formula is used [24][45]:

$$D(A, B) = \frac{\sum_{s=1}^{k} d(Ps) * Ws}{\sum_{s=1}^{k} Ws} \quad (1)$$

Where *D* is distance between sequences A and B; *d* (*Ps*) is distance between $i_s$ and $j_s$; *P* is the function representing points across the optimized (least distance) path between the two sequences; *Ws* > 0 is weighting coefficient. Detectors use the weighting coefficient such that

$$C = \sum_{s=1}^{k} Ws \quad (2)$$

Here *C* = *n* + *m* as we use the symmetric form for the weighting coefficient

$$D(A,B) = 1/C \sum_{s=1}^{k} d(Ps) * Ws \qquad (3)$$

The detectors calculate the optimized value for $D$ (A, B) using dynamic programming. Dynamic programming is a method for solving complex problems by breaking them down into simpler sub-problems [16] [45]. Dynamic programming is applied in the calculation as follows [45]:

Initial condition: $g$ (1, 1) = $2d$ (1, 1).

$$g(i,j) = min \begin{Bmatrix} g(i,j-1) + d(i,j) \\ g(i-1,j-1) + 2d(i,j) \\ g(i-1,j) + d(i,f) \end{Bmatrix} \qquad (4)$$

Where $g$ ($i$, $j$): min. is value of function *P* at point ($i$, $j$)

$$D (A, B) = g (n, m)/C \qquad (5)$$

Where $d$ ($i$, $j$) is Euclidian distance between point *i* of sequence1 and point *j* of sequence2.
Thus, if $D$ (A, B) <$D$ (B, C) then the sequences A and B are more similar as compared to sequences B and C [45]. The sequences mentioned in the above formulae could either be patterns in the pattern set of detectors or could be the stream of network traffic data encountered during the actual execution of the system or the data fed during the training phase.

In case of the Innate Immune System the value of *P* with which an intrusion is detected, is dependent on the value of D calculated using the DTW algorithm for the network traffic data stream. It is important to note that since Innate Immune System is based primarily on pattern recognition using the pattern sets for detectors, it is generic in nature i.e.it is not specific to any particular type of attack. As a result the primary response of an Innate Immune System is non-specific similar to the Innate Immune System in Human Immune System.

## 5. ADAPTIVE SYSTEM

Adaptive system is the most intelligent and computationally intensive component of the suggested system. As opposed to the Innate Immune System, Adaptive System does not work on the network traffic in its raw form. Data corresponding to the network traffic that pass through the Innate Immune System is first transformed to include only relevant feature sets in the data to be analyzed. The Adaptive System then analyses this transformed data based on mature functions generated during the training phase. These mature functions are generated using various machine learning techniques (both supervised and unsupervised data). Since Adaptive System works on transformed datasets and uses mature functions to analyze network traffic for intrusion, the responses initiated by this component correspond to secondary immune responses.

Adaptive system consists of two sub-components: Data Optimizer and Data Analyzer. We refer to the sub-component that transforms the network traffic data before it is fed to the Adaptive System as Data Optimizer. The details of Data Optimizer are as follows:

### 5.1 Data Optimizer

Network traffic data consists of multiple parameters like source, destination IP Addresses, incoming and outgoing data rate etc. It is important to note that in spite of the existence of the above mentioned parameters, the variability in the network traffic data is characterized by a limited feature set. This feature set characterizes the network traffic data and is instrumental in inferring patterns in the network traffic data. Thus projecting the data along these feature sets rather than using the raw data is more efficient. Data Optimizer thus performs feature selection [33] on the network traffic data before the network traffic data is passed to the Data Analyzer. The Data Optimizer applies Principal Component Analysis [34] to identify features that account for the variability in the training dataset and hence potential patterns that describe the training data set. Data Optimizer uses Principle Component Analysis primarily for dimension reduction, similar to the technique suggested in [35], which may be defined as the process of reducing the number of random variables under consideration in a given data set using feature selection and feature extraction techniques. During the training phase the Data Optimizer calculates the covariance matrix for the data set consisting of $n$ variable. Corresponding to this covariance matrix; the Data Optimizer records $n$ different eigenvectors and their corresponding eigenvalues [36]. Eigenvectors corresponding to $p$ highest eigenvalues are chosen as they account for the maximum variability of the network traffic data. Using these eigenvectors, a feature vector is calculated such that each column of the feature vector corresponds to one of the eigenvectors. The Data Optimizer transforms the network traffic data by projecting the original data along the $p$ new features. The motivation of the Data Optimizer behind using Principal Component Analysis is that it enables us to efficiently perform analysis on the transformed dataset for only those features that characterize the network traffic data.

**5.2 Data Analyzer**

The optimized network traffic data fed to the Data Analyzer by the Data Optimizer during the training phase is analyzed using various machine learning techniques. The concept of data dimensionality is used by the Data Analyzer. A data dimension may be defined as a data element that categorizes each item in a data set into non-overlapping regions [42].

Analysis of datasets along multiple dimensions plays an important role in Business Intelligence (BI). Analysis of business critical data along multiple dimensions reveals Key Performance Indicators (KPI) that are important for understanding various trends and patterns in businesses [27]. Multi-dimensional analysis of data yields significant advantages not only in businesses but also in various fields of study such as engineering, medicine, geology etc. [27][28]. The Data Analyzer incorporates the multi-dimensional view of network traffic data to analyze network traffic features for intrusion detection as follows:

Each of the features extracted by the Data Optimizer in the training phase is considered as different dimensions. For each of these dimensions the Data Analyzer infers a mature function using supervised learning techniques [3]. Apart from inferring mature dimension functions the Data Analyzer also generates a global mature $n$ variable function, where $n$ is the number of features. During intrusion detection for live traffic, the Data Analyzer applies the dimension values to each of the dimension functions to infer the existence of an intrusion in the system with respect to that dimension. The output of the individual dimension functions is then applied to the global function to determine the existence of an intrusion in the system by taking into consideration all the dimensions together.

In order to generate the mature dimension functions and the global mature function considering all the features together, Data Analyzer uses supervised learning techniques [3]. The primary reason for using supervised learning techniques is that network traffic data available is labeled data which is extracted using network traces from earlier well known attacks or publicly available datasets, like the KDD Cup 1999 Data [47]. The Data Analyzer primarily uses the following supervised learning techniques for generation of the mature functions:

1) Decision Tree
2) Sequential Learning

The primary reason for considering these two techniques is as follows:

1) Values for network traffic and system usage parameters and hence the derived features represent continuous and sequential labeled data, sequential learning techniques help to detect the label corresponding to an unknown observation based on the sequence encountered till the unknown observation [26][29].

2) Abnormality with respect to the features derived using Principal Component Analysis can be effectively expressed using *if then* rules. These *if then* rules can be effectively developed using a decision tree [25].
3) Regression trees not only help to infer *if then* rules but also help in determining potential dimensions that can be used for detecting the abnormality in system usage [24].

## 6. PROPAGATION OFINFORMATION TOOTHER INSTANCES OF THE SYSTEM

As explained earlier a typical network deployment of the proposed system involves multiple instances of the system operating on individual hosts of the network. These instances propagate system specific information to other instances of the system at regular intervals. The information that each of the instances is required to propagate consists of the following:

1) The values for malicious IP addresses and TCP ports that the instance has inferred (Surface Barrier).

2) The self and non-self patterns that the Innate Immune System of the instance has inferred.

3) The mature functions for the features inferred by the Adaptive System of the instances

In a distributed environment, the effective functioning of the system is dependent on the effective co-operation of the multiple instances of the system. Each instance of the system propagates information to other instances using the reputation based approach suggested in [6], wherein each instance will propagate the information to only its immediate neighboring instances. Individual components of the instances on receiving this information take the following actions:

1) The surface barrier updates it reference detection data set with new values for malicious IP addresses or TCP ports.

2) The Innate Immune System updates its self and non-self patterns based on the additional patterns inferred by other instances.

3) The Adaptive System improves its mature function set for each feature in the feature set based on the information received from other instances.

It is important to note that individual instances of the system must identify each other to avoid any incorrect or irrelevant information being fed to other instances of the system. Each instance uses a multi-factor authentication mechanism suggested in [41] to authenticate other instances and receive information from only authenticated instances. Apart from the technique suggested in [41], the instances also use a challenge-response based authentication system suggested in [40]. Security of the information exchanged between the instances of the system is ensured by transforming and fragmenting the data packets that carry the necessary information before transmitting them as suggested in [7]. An alternative technique used by the instances for secure information transfer is based on the concept of *jigsaw puzzle* suggested in [17]. Apart from fragmenting the information exchanged between multiple instances of the system, each of the instances hide the fragmented information in the TCP packets using either the LSB data hiding technique suggested in [46] or the steganography technique suggested in [44]. The rationale behind using the approaches suggested in [7] [17] is that, they ensure the confidentially and integrity of the transferred information. Thus no intermediate node or any unauthorized party can access the information being transferred to its completeness. The data fragments at each of the receiving instances are then combined and actions mentioned above are performed based on the received combined information. Instances of the suggested system deployed in a mobile network exchange system information with each other using mobile routing protocols like; Destination-Sequenced Distance-Vector (DSDV) and Dynamic Source Routing (DSR) [43].

## 7. CONCLUSION

In conclusion, the suggested hybrid approach based on Artificial Immune System and Soft Computing is instrumental in detecting intrusions and malicious activities in a given network. The three primary components of the system: surface barrier, innate immune system and adaptive system provide a layered defense mechanism, which evolves over multiple executions to combat new emerging attacks. Pattern matching performed by the Innate Immune System using Dynamic Time Warping provides efficient recognition of the self and non-self patterns in the network traffic data stream. The use of computationally intensive soft computing and machine learning techniques by the Adaptive System, provides additional advantage as far as analyzing complex network traffic data is concerned.